\documentclass[superscriptaddress,nofootinbib,twocolumn]{revtex4-1}

\usepackage[colorlinks]{hyperref}
\hypersetup{
     colorlinks   = true,
     citecolor    = red,
	linkcolor=blue
}

\usepackage{bbm}
\usepackage{amsfonts}
\usepackage{mathrsfs}

\usepackage{amsmath,amssymb}

\usepackage{epsfig}
\usepackage{graphicx}               
\usepackage{url}
\usepackage{hyperref}
\usepackage{float}
\usepackage{pstricks}
\usepackage{color}
\usepackage{multirow}

\makeatletter
\def\simgt{\mathrel{\lower2.5pt\vbox{\lineskip=0pt\baselineskip=0pt
           \hbox{$>$}\hbox{$\sim$}}}}
\def\simlt{\mathrel{\lower2.5pt\vbox{\lineskip=0pt\baselineskip=0pt
           \hbox{$<$}\hbox{$\sim$}}}}
\makeatother

\newcommand{\be}{\begin{equation}}
\newcommand{\ee}{\end{equation}}
\newcommand{\bea}{\begin{eqnarray}}
\newcommand{\eea}{\end{eqnarray}}
\newcommand{\beq}{\begin{eqnarray}}
\newcommand{\eeq}{\end{eqnarray}}

\newcommand{\no}{\nonumber}

\def\lsim{\mathrel{\rlap{\lower4pt\hbox{\hskip1pt$\sim$}}
     \raise1pt\hbox{$<$}}}         
\def\gsim{\mathrel{\rlap{\lower4pt\hbox{\hskip1pt$\sim$}}
     \raise1pt\hbox{$>$}}}         

\begin{document}

\title{Detecting Ultralight Bosonic Dark Matter via Absorption in Superconductors}

\author{Yonit Hochberg}
\author{Tongyan Lin}
\author{Kathryn M. Zurek}
\affiliation{Theoretical Physics Group, Lawrence Berkeley National Laboratory, Berkeley, CA 94720 \\ Berkeley Center for Theoretical Physics, University of California, Berkeley, CA 94720}

\begin{abstract}
Superconducting targets have recently been proposed for the direct detection of dark matter as light as a keV, via elastic scattering off conduction electrons in Cooper pairs.  Detecting such light dark matter requires sensitivity to energies as small as the superconducting gap of ${\cal O}({\rm meV})$.  Here we show that these same superconducting devices can detect much lighter DM, of meV to eV mass, via dark matter absorption on a conduction electron, followed by emission of an athermal phonon. We demonstrate the power of this setup for relic kinetically mixed hidden photons, pseudoscalars, and scalars, showing the reach can exceed current astrophysical and terrestrial constraints with only a moderate exposure.
\end{abstract}

\maketitle

\section{Introduction}
\label{sec:intro}
Since the first hints of dark matter (DM) nearly a hundred years ago, the search has been on to understand its nature. In the last thirty years, theoretical attention as well as experimental development has focused, to a large degree, on the Weakly Interacting Massive Particle (WIMP) paradigm.  An essential part of the experimental WIMP-hunting program is the direct detection of relic dark matter in the halo of the Milky Way.  Existing efforts, such as Refs.~\cite{Aprile:2012nq,Akerib:2015rjg,Agnese:2014aze}, have had immense success in constraining dark matter in the GeV$-$TeV mass range, and ton-scale detectors~\cite{Aprile:2015uzo,Akerib:2015cja} will improve substantially on current reach in the near future.

These experiments are, however, limited in their reach of light DM candidates due to their $\sim{\rm keV}$ energy thresholds, corresponding to the kinetic energy of a $\sim{\rm GeV}$ mass DM particle. Nonetheless, it is becoming increasingly clear, both theoretically and experimentally, that well-motivated and detectable DM candidates can be found with mass below these thresholds.  Examples include asymmetric DM from a hidden sector \cite{Kaplan:2009ag}, mirror DM \cite{Mohapatra:2000rk,Mohapatra:2001sx}, MeV-GeV mass DM \cite{Boehm:2003hm,Pospelov:2007mp,Hooper:2008im,Kumar:2009bw,Morrissey:2009ur,Cohen:2010kn}, and strongly interacting massive particles \cite{Hochberg:2014dra,Hochberg:2014kqa}.  
At even smaller DM masses, candidates include very weakly-coupled particles such as hidden photons, axions or axion-like particles, and scalars (see Ref.~\cite{Essig:2013lka} and references therein).

Detecting these lighter DM candidates in direct detection experiments is challenging due to the smaller DM kinetic energy available in the scattering, and, as the DM mass drops below the nucleus mass, by the kinematics of recoiling from a heavy target.  The maximum energy deposition by DM in an elastic scattering event off a target of mass $m_T$ is $q^2/(2 m_T)$ where the maximum momentum transfer is $q = 2 \mu_r v_X$, with $\mu_r$ the DM-target reduced mass and $v_X \sim 10^{-3}$ the DM velocity.  Thus 10~GeV mass DM can deposit at most a few keV on a nucleus, while MeV mass DM can deposit a mere meV of energy in such a scattering, well beneath nuclear recoil thresholds.

Instead, once the DM mass drops below the nucleus mass, electron targets are able to capture a larger fraction of the DM's kinetic energy. Electronic ionization in an atom~\cite{Essig:2011nj}, and excitation to the conduction band in a semiconductor~\cite{Essig:2011nj,Graham:2012su,Essig:2015cda,Lee:2015qva}, have both been proposed as concrete mechanisms to detect DM by electron recoils, and an analysis utilizing Xenon10 data has already been performed~\cite{Essig:2012yx}.  These approaches are, however, inherently limited by the energy gap for exciting an electron in the systems, typically in the $1-10$~eV range, which forbids access to DM lighter than $1-10$~MeV.

Thus new technology must be found to detect DM with sub-MeV mass.  Recently, a proposal was made to utilize a superconducting target as a means to detect DM $X$ as light as the warm dark matter limit, with mass $m_X\sim$~keV~\cite{Hochberg:2015pha,Hochberg:2015fth}. Such light dark matter carries little momentum, $|\vec q| \sim (m_X/{\rm keV})\,{\rm eV}$, and even less kinetic energy, $\omega \sim (m_X/{\rm keV})\,{\rm meV}$.  For detection of very light DM via scattering processes, the superconducting detectors carry three major advantages.  First, the gap in a superconductor (of ${\cal O}(0.3\;{\rm  meV})$ in a metal like aluminum) is much smaller than the (approximately eV or more) gap in semiconductors such as germanium and silicon.
Second, the electrons in a metal at zero temperature are Fermi-degenerate and have a velocity $v_F \sim 10^{-2}$ that exceeds the DM velocity.  Kinematically, this feature is crucial for being able to extract {\em all} the kinetic energy of the DM in the scattering process.  (While important for DM scattering, this second feature turns out to be unimportant for the DM absorption process which is the focus of this paper.)  Third, the small non-zero gap is essential to decoupling the signature electron recoils from lattice vibrations of the metal, essentially assisting in controlling the thermal noise.

The purpose of this paper is to show that superconductors are powerful not only as a means of detecting DM that scatters off electrons, but also for absorbing ultralight bosonic DM.  Here, ultralight refers to DM with mass in the meV to eV range.\footnote{In a separate publication, we explore absorption of DM via semiconductor targets, in the complementary eV to keV DM mass range~\cite{semiconductors}.}  In this mass range, beneath an eV, the density of DM particles exceeds their $({\rm wave length})^{-3}$, and the DM forms a coherent field.
Assuming that this field couples to electrons, a superconductor is then an excellent absorber of the DM, in the same way that superconductors and metals are excellent absorbers of electromagnetic fields. For instance, we find that a kg-day exposure on a superconducting target is sufficient to exceed the stellar constraints for a hidden photon whose mass is obtained via the Stuckelberg mechanism.

The outline of this paper is as follows.  In Section~\ref{ssec:phonon} we discuss how metals can be efficient absorbers of low mass particles. The process we consider involves absorbing all the mass-energy of the DM particle via an electron recoil, with emission of an athermal phonon to conserve momentum.  We then describe in Sections~\ref{ssec:abs} and~\ref{ssec:absphoton} our method to determine the DM absorption rate from the optical properties of a metal. In Section~\ref{sec:app} we present the reach of superconducting detectors for ultralight DM that couples to electrons, including hidden photons, pseudoscalars, and scalars.
We conclude in Section~\ref{sec:conc}.

\section{Dark Matter Absorption with Superconductors}\label{sec:opt}

We begin by describing the DM absorption process, before computing its rate in a superconductor.  We compare our results for consistency against the standard Drude theory for low-energy photon absorption in metals. Then, in order to obtain accurate predictions at higher ($\gtrsim$ 0.1 eV) energies, we relate the DM absorption rate to measured photon absorption rates.

\subsection{General Principle: Phonon emission \label{ssec:phonon} }

Absorption of low energy particles in a superconductor can proceed when the energy of the absorbed radiation (in this case the mass of the DM particle) exceeds the superconducting gap.  In the absorption process, a Cooper pair is broken, and a pair of excitations is created.  These excitations have a long recombination and thermalization time (of order a few milliseconds in aluminum), which allows for their collection and measurement, as described in Refs.~\cite{Hochberg:2015pha,Hochberg:2015fth}.  Once the energy of the absorbed particle significantly exceeds the superconducting gap, the absorption process is identical in the superconducting and normal phases of a metal.  There are several ways to absorb a particle (be it a photon or DM) in a metal.  One way is via impurities, where an off-shell electron produced in the absorption process becomes on-shell through interaction with an impurity.  In the case of interest here, however, the target superconductor must be ultrapure in order to enable the collection and measurement of the created athermal excitations, and so this possibility is not viable.

Instead, we make use of another process -- that of particle absorption on electrons through the emission of an athermal phonon in the final state, as shown in Fig.~\ref{fig:feyn}.  The emitted phonon is required for momentum conservation of the target material.  Consider an electron with initial momentum $\vec k_i$ and energy $E_i=\vec k_i^2/(2 m_e)$. Assuming the electron absorbs a single particle of energy $ \omega$, the final momentum of the electron is  $\vec k_f = \vec k_i + \vec q$ and energy conservation gives
\begin{equation}
	\frac{ (\vec k_i + \vec q)^2}{2m_e} = \frac{ \vec k_i^2}{2m_e} + \omega.
	\label{eq:kinematics}
\end{equation}
(Note that momentum on the lattice is conserved up to an additive reciprocal lattice vector, $\vec K$. For electrons, the typical energy scale associated with transitions involving $\vec K$ is $K^2/2m_e \sim 10$ eV, which is above the energies considered here.)
Then the required momentum transfer to the electron is $|\vec q| \sim \omega (m_e/ |\vec k_i|) \sim \omega/v_F \sim 100\ \omega$, where $v_F$ is the Fermi velocity. This cannot be satisfied for an on-shell DM particle in the halo, which carries momentum $\sim 10^{-3} \omega$.
However, energy and momentum {\emph {can}} still be conserved if a phonon with momentum $\sim - \vec q$ is emitted by the electron in the final state;  in other words, the electron recoils against the lattice. The emitted phonon carries away a fraction of the excitation energy, but can balance the large recoil momentum of the electron. 

In the Debye model, the dispersion relation of a phonon with 4-momentum $(\Omega, \vec Q)$ is given by
\begin{equation}
		\Omega = c_s |\vec Q|
\end{equation}
where the speed of sound in aluminum is $c_s \simeq 6320\; {\rm m/sec}\sim 2\times 10^{-5}$ in natural units. There is a maximum frequency $\omega_D = c_s k_D$ for phonons, where the maximum wavevector for lattice vibrations $k_D \sim 1/a$ is set by the lattice spacing $a$. For aluminum, $\omega_D \approx$ 0.037~eV; therefore the maximum phonon energy is relatively low, but the maximum momentum can be much higher, $\omega_{D}/c_s \approx$~keV.

\subsection{Dark Matter Absorption}\label{ssec:abs}
%
\begin{figure}[t!]
\begin{center}
	\includegraphics[width=0.47\textwidth]{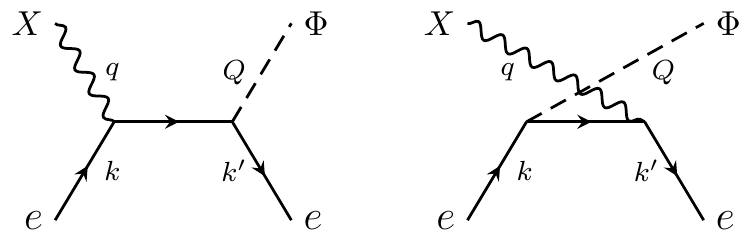}
\end{center}
 \caption{Absorption process on electrons for an incoming relic particle $X$, where a phonon $\Phi$ is emitted in the final state:  $X(q) + e(k) \to e(k')+\Phi(Q)$.
  \label{fig:feyn}}
\end{figure}

We now turn to computing the rate of DM absorption in a material. The total DM absorption rate per unit mass per unit time $R$ is
\beq
	\label{eq:R}
	R=  \frac{1}{\rho} \frac{\rho_X}{m_X}   \langle n_e\sigma_{\rm abs} v_{\rm rel}\rangle \,,
\eeq
where $\sigma_{\rm abs}$ is the absorption cross section on electrons, $\rho$ is the mass density of the target material, and $\rho_X=0.3 \; {\rm GeV}/{\rm cm}^3$ is the local mass density of DM.

Treating the target as a free electron gas with Fermi energy $E_F$, the rate for the $2\to 2$ process of $X(q) + e(k) \to e(k')+\Phi(Q)$ (with $\Phi$ a phonon) is given by
\begin{align}
	 \label{eq:response}
	\langle n_e\sigma_{\rm abs} v_{\rm rel}\rangle&=\int\frac{d^3 Q}{(2\pi)^3}\frac{\langle |{\cal M}|^2\rangle}{16 E_1 E_2 E_3 E_4}\ S(q, Q)\, , \\
	S(q,Q)&=2\int\frac{d^3 k}{(2\pi)^3}\frac{d^3 k'}{(2\pi)^3}(2\pi)^4\delta^4(k+q-k'-Q) \nonumber \\
	& \hspace{2cm} \times f(E)(1-f(E'))\,, \nonumber
\end{align}
where $\langle |{\cal M}|^2\rangle$ is the averaged and summed matrix-element-squared for the process. The functions $f(E)$ are electron occupation numbers, with $(1 - f(E'))$ characterizing Pauli blocking effects. The four-momentum of the absorbed particle is $q = (\omega, \vec q)$, while the emitted phonon has $Q = (\Omega, \vec Q)$ with $\Omega = c_s |\vec Q|$. For $T=0$ and $|\vec q|\ll \omega \ll E_F$, the integral over the initial and final electron phase space $S(q,Q)\approx S(\omega, \vec Q)$ reduces to a simple Heaviside theta function of allowed kinematic configuration, with amplitude
\begin{equation}
	S(\omega, \vec Q)\simeq (m_e^*)^2(\omega - \Omega)/(\pi |\vec Q|) \, .
\end{equation}
Here $m_e^*$ is the effective electron mass in the metal.

For each of the DM models we consider in Sec.~\ref{sec:app}, we compute $\langle |{\cal M}|^2\rangle$ for DM absorption via phonon-emission, treating the phonon as a scalar field $\Phi$ and assigning the electron-electron-phonon vertex with the dimensionless coupling
\beq
	y_{\Phi} = C_{\Phi}|\vec Q|/\sqrt{\rho}
	\label{PhononInt}
\eeq
(we refer the reader to Appendix J of Ref.~\cite{kittel} for a derivation of this result). The parameter $C_{\Phi}$ has units of energy and is of order $E_F$, but must be determined by matching onto data. 

In order to check the validity of this procedure and to fix the electron-phonon coupling using existing data, we must turn to \emph{photon} absorption. Photon absorption proceeds by a similar 2-to-2 process as DM absorption, and has been measured in aluminum over a range of energies. By comparing the data with the photon absorption rate computed with Eq.~\eqref{eq:response}, we can then obtain the coupling constant $C_{\Phi}$. Equivalently, we will find that the DM absorption rate can be written in terms of the photon absorption rate, and this relation holds even at larger $\omega$, where the free-electron approximation breaks down. We note that although the spatial momenta $|\vec q|$ of massive DM differs from that of the photon, this difference is unimportant for the absorption process. The reason is that the momentum of both the absorbed photon and DM particle is negligible compared to the electron momenta.

We first calculate the rate for photon absorption at low energies. Summing over the diagrams shown in Fig.~\ref{fig:feyn}, and averaging over incoming electron spin and photon polarizations, we find the matrix-element-squared in the limit of $\omega \ll |\vec Q|$ is given by
\begin{align}
	 |{\cal M}_\gamma|^2 \approx \frac{4 e^2}{3} \frac{C_{\Phi}^2 }{\rho} \frac{ |\vec Q|^4 }{ \omega^2} \, .
	\label{eq:Msquared_photon}
\end{align}
The total rate for photon absorption is then (for $\omega \ll E_F$, where $E_F=11.7$~eV in aluminum)
\begin{align}
	 \langle n_e\sigma_{{\rm abs}} v_{\rm rel}\rangle_\gamma & \simeq \frac{n_e e^2}{m_e^*\, \omega^2} \left( \frac{2\pi}{\omega} \int \frac{d\Omega (\omega - \Omega) \Omega^4 }{3\,(2\pi)^4 } \frac{  C_{\Phi}^2}{ c_s^6\rho} \frac{m_e^*}{n_e} \right)   \nonumber \\
	& \equiv \frac{n_e e^2}{m_e^* \omega^2} \frac{1}{\tau(\omega)} \, .
	\label{eq:rateQFT}
\end{align}
The integral over $\Omega$ is restricted to energies either below $\omega$ (due to energy conservation) or below $\omega_D$ (due to the cutoff in phonon momenta), whichever is smaller.   Above we have suggestively defined the $\omega$-dependent parameter $\tau(\omega)$ as the quantity in parenthesis in the first line of Eq.~\eqref{eq:rateQFT}, in order to compare this result to the standard theory for absorption of EM fields in metals, the Drude theory.  We will see next that $\tau(\omega)$ is a time-scale for phonon emission.

\subsection{Photon Absorption and Superconductor Response}\label{ssec:absphoton}

In order to make a connection between our calculation of the photon absorption rate, Eq.~(\ref{eq:rateQFT}), and the Drude theory, we begin by noting that the absorption rate of photons can be related to the polarization tensor of the EM field $\Pi$ via the optical theorem:
\beq
	\langle n_e\sigma_{{\rm abs}} v_{\rm rel}\rangle_\gamma = -\frac{{\rm Im}\;\Pi(\omega)}{\omega}\,,
\eeq
where in the local limit of $|\vec q| \ll \omega$ the transverse and longitudinal modes of the polarization tensor are of equal size, which we denote by $\Pi(\omega)$. This $\Pi$ is related to the complex conductivity $\hat \sigma(\omega) \equiv \sigma_1+i\sigma_2$, describing the frequency-dependent response of electrons to an EM perturbation, by
\beq\label{eq:pisig}
	\Pi(\omega)\approx - i \hat \sigma \omega\,.
\eeq
(See Appendix~\ref{app:Pi} and {\it e.g.} Ref.~\cite{Hochberg:2015fth} for further details.) As is evident, the real part of the conductivity $\sigma_1$ is the absorption rate for excitations of energy $\omega$, and is related to the absorption cross section of photons by
\beq\label{eq:sigabs}
	\sigma_1 = \langle n_e \sigma_{{\rm abs}} v_{\rm rel} \rangle_\gamma \, ,
\eeq
making clear from Eq.~\eqref{eq:R} that large non-zero $\sigma_1$ is crucial for absorption.

We can now compare the rate in Eq.~(\ref{eq:rateQFT}) to the conductivity derived from the Drude model.  The Drude model describes the conductivity at energies above the superconducting gap and below the gap for direct transitions between bands ($\sim$~eV)~\cite{dresselgruner}. The Drude theory gives
\begin{equation}
	\hat \sigma(\omega) = \frac{n_e e^2 \tau}{m_e^*} \frac{1}{1 - i \omega \tau}\,,
\end{equation}
with real and imaginary parts
\beq\label{eq:sig1}
	\sigma_1(\omega) &=& \omega_p^2 \tau \frac{1}{1 + \omega^2 \tau^2} \approx \frac{\omega_p^2}{\omega^2 \tau}  \,, \\
	\sigma_2(\omega) &=& \omega_p^2 \tau \frac{\omega \tau}{1 + \omega^2 \tau^2} \approx \frac{\omega_p^2}{\omega} \,,\label{eq:sig2}
\eeq
where the last approximation is in the $\omega \tau \gg 1$ limit, and the plasma frequency $\omega_p$ is
\begin{equation}\label{eq:omegap}
			\omega_p = \left( \frac{n_e e^2}{m_e^*} \right)^{1/2}\,.
\end{equation}
 We immediately see the correspondence between $\sigma_1$ in the Drude theory and the result Eq.~(\ref{eq:rateQFT}), once $\tau$ is determined in the Drude theory. 
In what follows we use $\omega_p = 12.2$ eV for aluminum~\cite{handbook}. 

The parameter $\tau$ represents an electron scattering time in the medium. In general, $\tau$ is both temperature and $\omega$-dependent. In the $\omega \to 0$ limit and at low temperatures, $\tau$ is primarily set by the impurities of the system and determines the DC conductivity. However, in the $ \omega \gg T$ limit relevant to us (the operating temperatures of the proposed superconducting detectors are ${\cal O}(10{\rm mK})\sim \mu{\rm eV}$), $\tau$ is set by electron interactions with athermal phonons. Using the simple Debye model for the phonon dispersion, the phonon-electron interactions give $\tau=\tau_{\Phi}$, where the rate for the electron to emit the phonon is~\cite{PhysRevB.3.305}
\begin{align}
	\frac{1}{\tau_{\Phi}} = \begin{cases}
		\frac{4}{5} \pi \lambda_{\rm tr} \omega_D \left( 1 - \frac{5}{6} \frac{\omega_D}{\omega} \right) & , \ \omega \ge \omega_D \\
		\frac{2}{15} \pi \lambda_{\rm tr} \frac{\omega^5}{\omega_D^4} & ,\ \omega < \omega_D
	\end{cases}.
	\label{eq:tauphonon}
\end{align}
For aluminum, $\omega_D \approx 0.037$ eV, and the measured high-temperature resistivity gives $\lambda_{\rm tr} = 0.39$~\cite{PhysRevB.36.2920}. Then $\omega\tau\gg1$ and we see that Eq.~(\ref{eq:rateQFT}) thus gives the same result as the Drude model, Eq.~(\ref{eq:sig1}), which can be used to fix $C_\Phi$. In practice we will use the Drude model, normalizing $\tau_\Phi$ by comparing directly with data.

\begin{figure}[t!]
\begin{center}
	\includegraphics[width=0.48\textwidth]{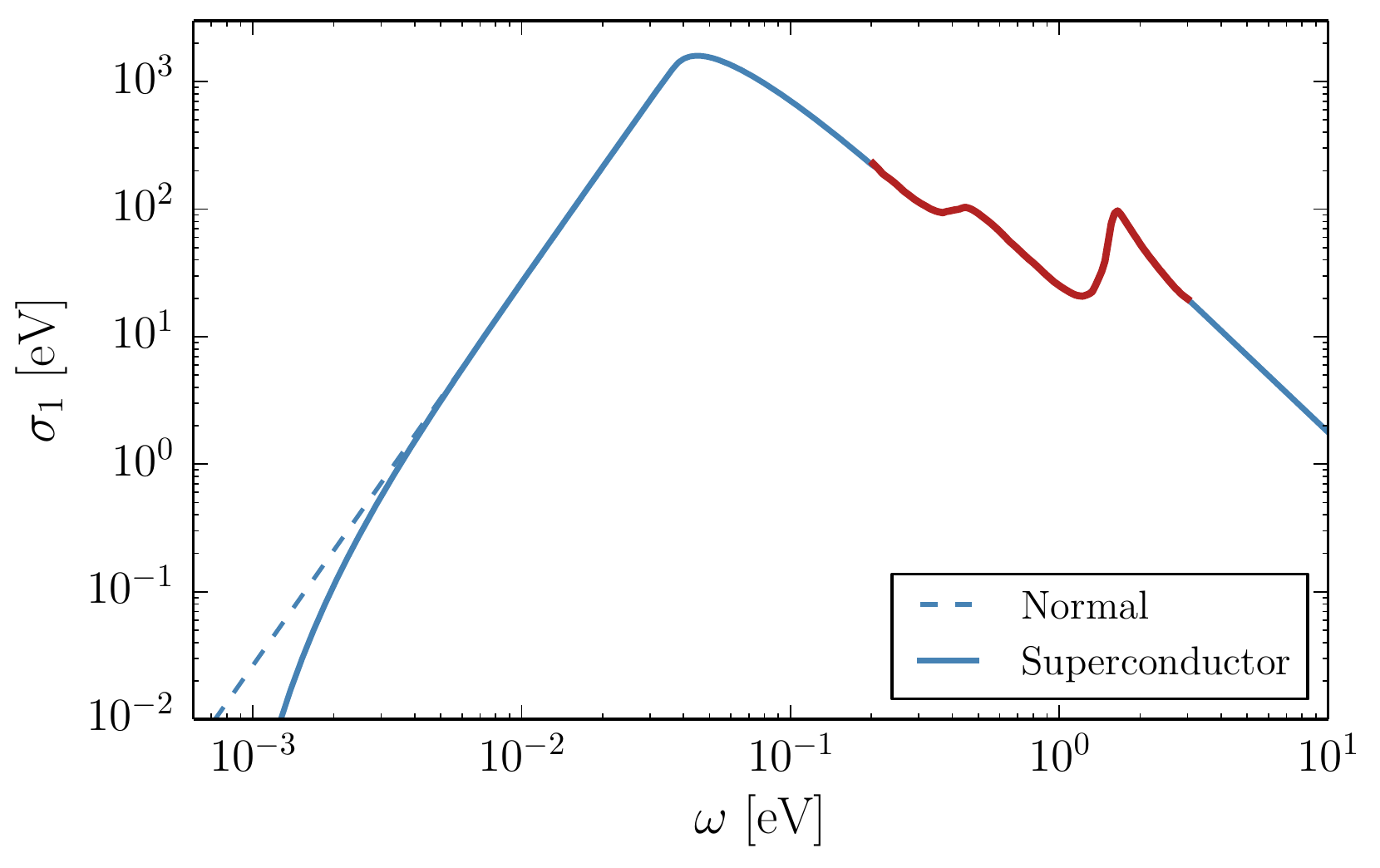}
\end{center}
 \caption{Absorptive part of conductivity in low temperature aluminum: below $\omega = 0.2$ eV, we use the analytic Drude theory, Eqs.~\eqref{eq:sig1} and~\eqref{eq:tauphonon}, here shown in the normal metal phase (dashed blue curve) and with the inclusion of coherence effects (Eq.~\eqref{eq:tauS}) in the superconducting phase (solid blue curve); we match this onto low-temperature data~\cite{PhysRevB.12.5615} (solid thick red curve); and then extrapolate to higher energies.
  \label{fig:sigdata}}
\end{figure}

The Drude theory, from a strict point of view, applies only for a metal in the normal (non-superconducting) phase. For the case of absorption, however, the difference between an ordinary metal and a superconductor is only relevant when the absorbed particle energy is close to twice the superconducting gap, $2\Delta$, which is the minimum energy required to break a Cooper pair.  Once the absorbed energy is much larger than $2\Delta$, the system is once again described by the free electron model of a metal. Near the gap, the modification of the absorption rate in a superconductor relative to that of a metal can be encoded in a so-called coherence factor. Following Ref.~\cite{PhysRevB.3.305}, we include this effect on the rate by using a different $\tau_\Phi^S$ in the superconducting phase, which is related to the normal metal phase $\tau_\Phi^N$ close to the gap by
\begin{align}\label{eq:tauS}
	\frac{ \tau_\Phi^N}{\tau_\Phi^S} = \frac{ \int^{\omega-2\Delta}_0 d\Omega (\omega - \Omega) \Omega^4  \textrm{ \large $E$}  \left[  (1 - \frac{4\Delta^2}{(\omega - \Omega)^2})^{1/2} \right] }{  \int^{\omega}_0 d\Omega (\omega - \Omega) \Omega^4 },
\end{align}
where $E$ is the complete elliptic integral of the second kind.
The inclusion of this factor only modestly affects our results near threshold.

\begin{figure*}[t!]
\begin{center}
	\includegraphics[width=0.68\textwidth]{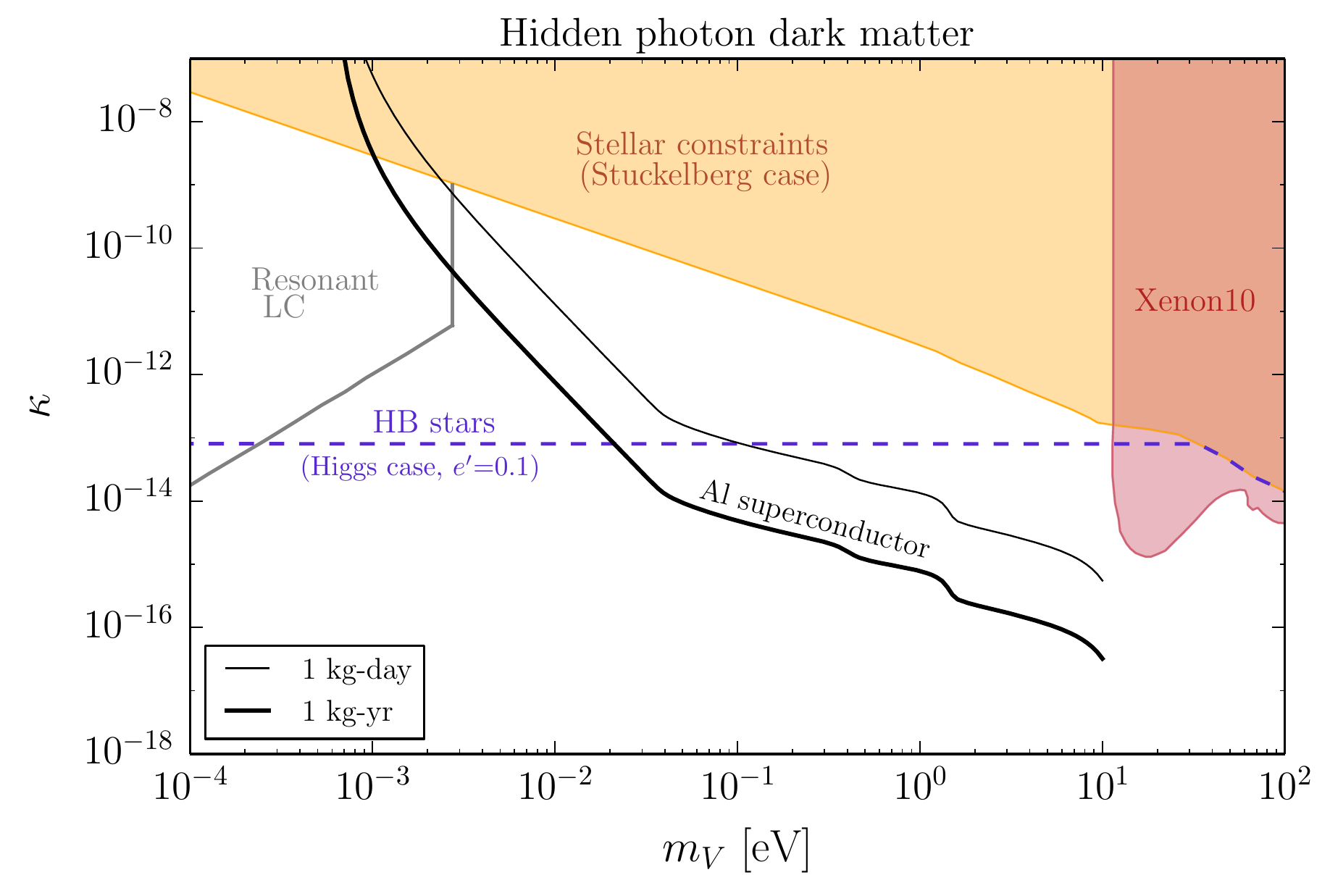}
\end{center}
 \caption{
Estimated sensitivity of an aluminum superconductor target for 1-kg-year (thick solid black) and 1-kg-day (thin solid black) exposures, for absorption of hidden photon relic dark matter. For comparison, we show solar and horizontal branch constraints for the Stuckelberg (shaded orange) and Higgs cases (dashed purple)~\cite{An:2013b}; Xenon10 bounds (shaded red)~\cite{An:2014twa}; and the projected reach for an LC circuit experiment (solid gray curve)~\cite{Chaudhuri:2014dla}.
  \label{fig:sensitivityV}}
\end{figure*}

For higher energies ($\omega \sim 0.5$ eV in aluminum), inter-band transitions are possible, and the Drude theory is incomplete. In principle, the integral in Eq.~\eqref{eq:response} over electron momentum states must be modified to take into account the full band structure of the material. Fortunately, measurements of photon absorption in aluminum are available in this energy range and, where possible, we directly obtain $\sigma_1$ from the data. As long as we can simply relate the matrix-element-squared of DM absorption to photon absorption, we are free to use measured $\sigma_1$ to normalize absorption rates, rather than performing the many-body calculation.

To summarize, we determine $\sigma_1$ over the meV-10 eV energy range through a combination of theoretical calculation and experimental measurements.
Our resulting $\sigma_1$ for aluminum is shown in Fig.~\ref{fig:sigdata}. At the lowest energies, we use the analytic result in the Drude theory, Eqs.~\eqref{eq:sig1} and~\eqref{eq:tauphonon}, including the coherence effects close to the superconducting gap at $\Delta \simeq$ 0.3 meV using Eq.~\eqref{eq:tauS}. We fix the overall normalization of $\tau_\Phi$ by matching onto low-temperature data on $\sigma_1$ in the 0.2--3 eV energy range~\cite{PhysRevB.12.5615}.
From 3 eV to 10 eV, we extrapolate $\sigma_1$ with an $\omega^{-3}$ power law; we note that the slope and amplitude of this power law closely follows $\sigma_1$ measured at room temperature~\cite{hagemannOSA,hagemannDESY}.

In what follows, we use the results of this section to relate the DM absorption rate to that of a photon, and then apply the combined solid $\sigma_1$ curve of Fig.~\ref{fig:sigdata} to derive the sensitivity of a superconducting aluminum target to various DM candidates.

For the hidden photon model described next, we will also require knowledge of $\sigma_2$ at low temperatures; here we simply use the result in the Drude theory, Eq.~\eqref{eq:sig2}, over the whole energy range. We have verified the validity of this approximation by comparing with measurements of $\sigma_2$ at room temperature~\cite{hagemannOSA,hagemannDESY}, finding at most $\sim 50\%$ difference with the Drude theory.

\section{Rates and Constraints}\label{sec:app}

Utilizing the results of the previous section, we now turn to ultralight bosonic DM~--- hidden photons, pseudoscalars, and scalars~--- in each case assuming that the candidate composes all the DM.

\subsection{Dark Photons}

Consider a hidden photon which is kinetically mixed with the hypercharge gauge boson, leading to kinetic mixing with the photon,
\begin{equation}
	{\cal L}\supset -\frac{\kappa}{2} F_{\mu \nu} V^{\mu \nu}\,,
\end{equation}
where $F^{\,u\nu}\;(V^{\mu\nu})$ are the field strengths for the photon (hidden photon).  For the parameter space considered here, this hidden photon may be all of the DM, where the origin of the relic abundance is set by a misalignment mechanism during or before inflation~\cite{Nelson:2011sf,Arias:2012az,Graham:2015rva}.

Performing a field redefinition of the photon $A_\mu \to A_\mu - \kappa V_\mu$ leads to the canonical basis, where the electromagnetic current $J_{\rm EM}^\mu$ picks up a dark charge, $\kappa e V_\mu J_{\rm EM}^\mu$ in vacuum. However, this mixing angle can vary substantially from $\kappa$ due to in-medium effects, which affect the polarization tensor $\Pi$ (related to the conductivity $\hat\sigma$ via Eq.~\eqref{eq:pisig}). In a metallic target such as aluminum, the effective mixing angle is suppressed by powers of the plasma frequency,
\begin{align}
	\kappa^2_{\rm eff} =  \frac{ \kappa^2 m_V^4}{ \left[ m_V^2 - {\rm Re}\;\Pi(\omega) \right]^2 + \left[ {\rm Im}\;\Pi(\omega) \right]^2} \simeq  \frac{ \kappa^2 m_V^4}{\omega_p^4}\,,
\end{align}
where we used Eqs.~\eqref{eq:pisig},~\eqref{eq:sig1}, and \eqref{eq:sig2}. Since ${\rm Re}\,\Pi \approx \omega \sigma_2 $ is larger than both ${\rm Im}\,\Pi \approx \omega \sigma_1$ and $m_V^2 \simeq \omega^2$ in our region of interest, we then have $\kappa_{\rm eff} \ll \kappa$.  Note that the suppression by the plasma frequency is different than the electron-scattering case explored in Refs.~\cite{Hochberg:2015pha,Hochberg:2015fth}, where the Thomas-Fermi screening length was relevant for determining $\kappa_{\rm eff}$.  The reason is that the absorption process occurs when the momentum transfer is much smaller than the absorbed energy, $|\vec q| \sim 10^{-3} m \ll \omega$, whereas scattering in the non-relativistic limit occurs when $|\vec q| \gg \omega$. (See Sec.~5.2 of Ref.~\cite{Hochberg:2015fth} for a discussion of the $(q,\omega)$-dependence of the screening mass.)

\begin{figure*}[t!]
\begin{center}
	\includegraphics[width=0.68\textwidth]{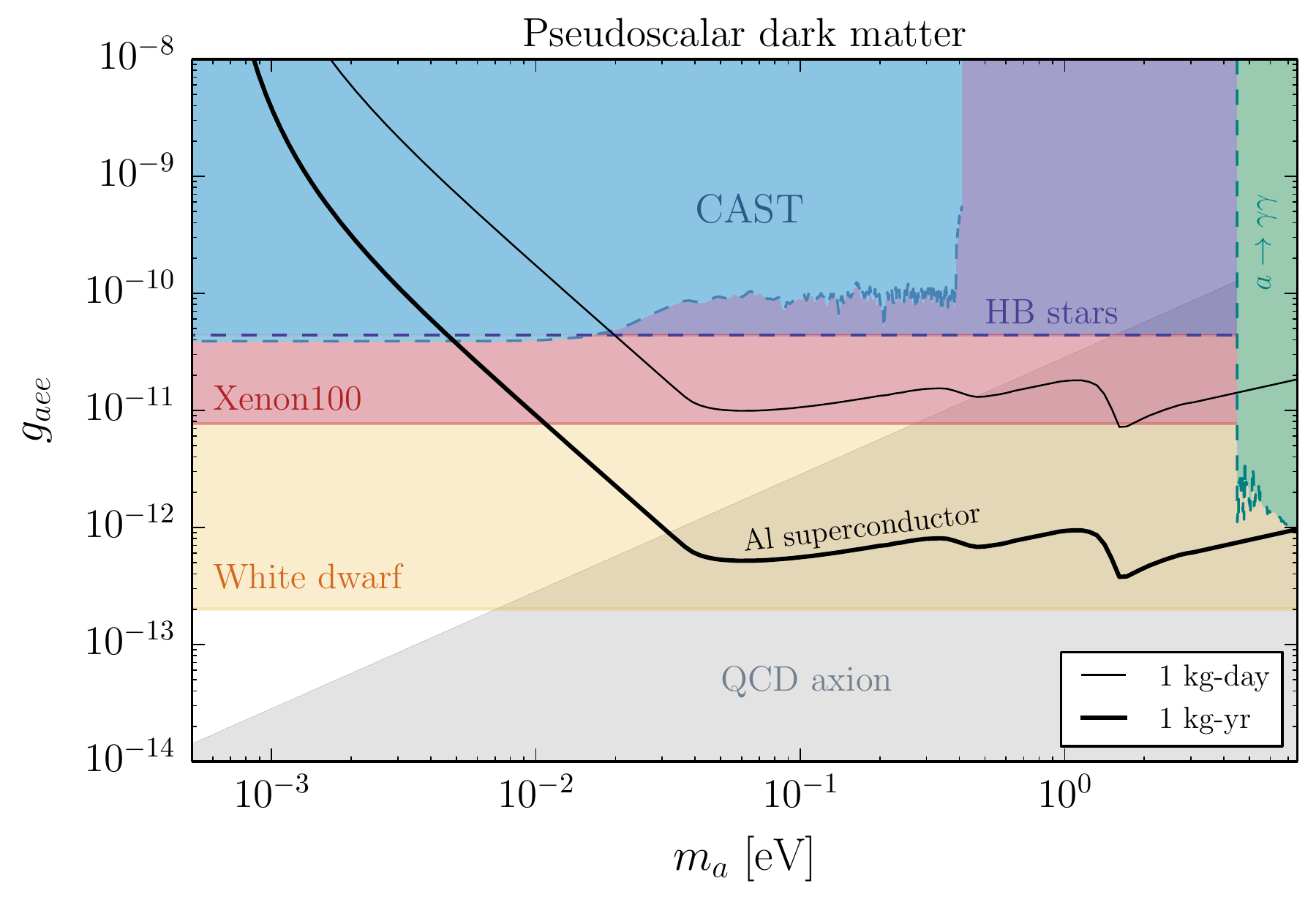}
\end{center}
 \caption{Estimated sensitivity of an aluminum superconductor target for 1-kg-year (thick solid black) and 1-kg-day (thin solid black) exposures, for absorption of pseudoscalar relic dark matter. For comparison, we also show constraints from absorption of solar axions in Xenon100 (shaded pink)~\cite{Aprile:2014eoa}; stellar emission from white dwarfs (shaded orange)~\cite{Raffelt:2006cw}; as well as the QCD axion relation (shaded gray). Dashed lines show constraints from a loop-induced {\emph{photon}} coupling given by Eq.~\eqref{eq:axion_photon}, which assumes the pseudoscalar does not couple to other charged particles. Such constraints include emission from horizontal branch (HB) stars (shaded purple)~\cite{Raffelt:2006cw}; the CAST experiment (shaded blue)~\cite{Arik:2008mq}; and decays into photons (shaded green)~\cite{Grin:2006aw}; these constraints are taken from studies that assume only a photon-coupling.
  \label{fig:sensitivityA}}
\end{figure*}

For the absorption of the kinetically mixed hidden photon, the matrix-element-squared is simply related to that of the photon by $|{\cal M}|^2  =   \kappa_{\rm eff}^2 |{\cal M}_\gamma|^2$.
Then the rate in counts per unit time per unit target mass, Eq.~\eqref{eq:R}, is found to be
\beq
	R=  \frac{1}{\rho}  \frac{\rho_{\rm DM}}{m_{\rm DM}}  \kappa_{\rm eff}^2  \sigma_1 \,.
\eeq
The projected sensitivity for a hidden photon is presented in Fig.~\ref{fig:sensitivityV}, assuming the particle comprises all of the DM and a kg-day (thin solid black curve) or kg-year (thick solid black curve) exposure.  Considering energy depositions between 1~meV to 1~eV, this corresponds to 3.6 events at 95\% CL, since the solar neutrino background is expected to produce fewer than an event in a kg$\cdot$year~\cite{Hochberg:2015fth}; since the absorption signal is mono-energetic, we assume for simplicity no background for DM masses in the eV to 10~eV energy range. Note also that a higher energy threshold for the experiment would correspond simply to cutting off the reach at lower DM masses, leaving the high-mass region unaffected.

Direct detection constraints on relic vector DM via an absorption process have been derived in Ref.~\cite{An:2014twa,Pospelov:2008jk} for masses above 12~eV, using low-threshold Xenon10 data (depicted in Fig.~\ref{fig:sensitivityV} in shaded red), and above 1~keV, using Xenon100 data. For masses in these ranges, absorption on semiconductor targets such as germanium and silicon should be competitive, and will be presented elsewhere~\cite{semiconductors}. Constraints from stellar emission in the sun and horizontal branch (HB) stars on masses below a few 10's of keV are relevant as well~\cite{An:2013a,An:2013b}, and are shown in Fig.~\ref{fig:sensitivityV}. The dominant emission process varies depending on whether a dark Higgs boson is present in the theory or not. In the former case, the bounds depend on the charge of the dark Higgs under a dark $U(1)$ (denoted $e'$, with $e'\,\kappa$ constrained), while in the latter case there is no such dependence; see Refs.~\cite{An:2013a,An:2013b} for details. These constraints are depicted in Fig.~\ref{fig:sensitivityV}, marked as `Higgs' (dashed purple) and `Stuckelberg' (shaded orange) accordingly.

A recent proposal to detect the hidden photon field with resonant LC circuits~\cite{Chaudhuri:2014dla} estimates strong sensitivity below 3~meV (and extending as far down as $10^{-12}$ eV). These projections are depicted by the gray solid curve in Fig.~\ref{fig:sensitivityV}. A  multiplexed version of this experiment could potentially reach mixings of $\kappa \sim 10^{-16}$ for meV masses.

We learn that an aluminum superconductor target with a kg-year exposure can be more sensitive than stellar constraints over the entire mass range of interest, from 1~meV to 10~eV, if the hidden photon obtains its mass via a Stuckelberg mechanism. If a dark Higgs is present, superconducting targets  with a kg-year exposure are stronger probes than horizontal branch stars for vector masses heavier than about 20~meV, for $e'\sim 0.1$. Since stellar emission depends on the stellar environment and as such is model-dependent, direct detection provides a strong orthogonal probe to such constraints.

\begin{figure*}[!t]
\begin{center}
	\includegraphics[width=0.69\textwidth]{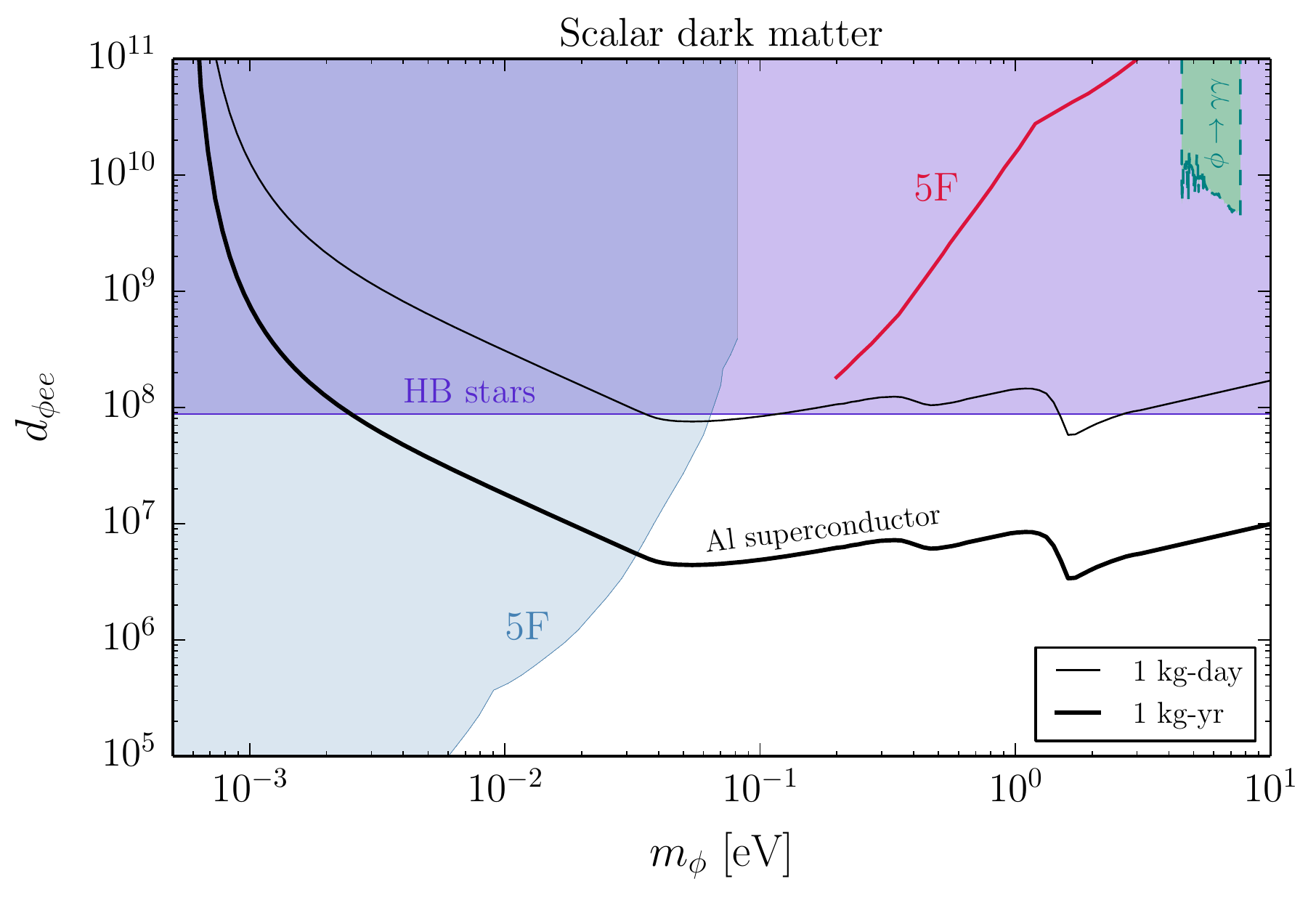}
\end{center}
 \caption{Estimated sensitivity of an aluminum superconductor target for 1-kg-year (thick solid black) and 1-kg-day (thin solid black) exposures, for absorption of scalar dark matter. For comparison, we also show constraints from fifth-force (shaded blue and solid red curve)~\cite{Adelberger:2003zx}; horizontal branch (HB) cooling (shaded purple)~\cite{Raffelt:1996wa,Grifols:1988fv}; and decays into photons (dashed green outline)~\cite{Grin:2006aw}.
  \label{fig:scalar}}
\end{figure*}

\subsection{Pseudoscalars}

We now proceed to pseudoscalars $X=a$ coupling to electrons:
\begin{equation}
	{\cal L}\supset \frac{g_{aee}}{2 m_e} (\partial_\mu a)\bar e \gamma^\mu \gamma^5 e\,.
\end{equation}
While a candidate for $a$ is the QCD axion, the relic density for the QCD axion cannot saturate the observed DM relic abundance in the mass range we consider, at least in the standard cosmology. More exotic mechanisms may be required for QCD axions to be all of the DM; alternatively, the pseudoscalar may be an axion-like particle~\cite{Arias:2012az}.

Comparing the pseudoscalar matrix-element-squared to the case of a photon, we find the same leading $\vec Q$-dependence,
$|{\cal M}|^2  \approx  3 ( g_{aee}/2 m_e)^2 (\omega/e)^2 |{\cal M}_\gamma|^2$.
Then the DM absorption rate is related to the conductivity as:
\begin{equation}
\label{eq:rateAxion}
	R = \frac{1}{\rho} \frac{\rho_X}{m_X}  \frac{3 m_a^2}{4 m_e^2}  \frac{g_{aee}^2}{e^2}   \sigma_1 \,.
\end{equation}

The expected reach into the parameter space of pseudoscalar DM via absorption on an aluminum superconducting target is shown in Fig.~\ref{fig:sensitivityA}, for a kg-day (thin solid black curve) and kg-year (thick solid black curve) exposure.  Stellar constraints on light pseudoscalars are shown as well~--- the electron coupling allows for emission of the pseudoscalar in the mass range of interest within electron-dense environments such as white dwarfs.  The cooling curves of white dwarfs give the strongest constraints on the electron coupling over our entire mass range~\cite{Raffelt:2006cw}. It has been argued that some of the data are in favor of a new weakly coupled particle~\cite{Giannotti:2015kwo}, and the limits shown are subject to a factor of a few uncertainty. We also show constraints from Xenon100~\cite{Aprile:2014eoa} (shaded pink) on DM emitted from the sun, which have keV energies and can be detected via an absorption process.

For completeness, we also show the relation between mass and $f_a$ for the QCD axion, $(0.60\ {\rm meV}/ m_a) = (f_a/10^{10}\ {\rm GeV} )$. Then the effective coupling can be written as $g_{aee} = C_e m_e/f_a$, where for DFSZ axions, $C_e = \frac{1}{3} \cos^2 \beta$, and for KSVZ axions with only a loop-induced electron-coupling, $C_e \propto \alpha^2$. In the shaded grey region, we take as an upper bound $C_e = 1/3$. 

Given an electron coupling, a loop-induced coupling of the pseudoscalar to photons arises,
\begin{equation}
	\frac{\alpha}{8\pi } \frac{g_{aee}}{m_e} a F_{\mu \nu} \tilde F^{\mu \nu}\,.
	\label{eq:axion_photon}
\end{equation}
If the pseudoscalar couples to other charged particles, this coupling will be modified by an ${\cal O}(1)$ factor.  Assuming only the induced photon coupling above, we can place constraints on $g_{aee}$ from CAST~\cite{Arik:2008mq} (shaded blue), cooling of HB stars (shaded purple), and the $a \to \gamma \gamma$ decay time~\cite{Grin:2006aw} (shaded green). (The IAXO experiment is expected to improve on the constraint from CAST by at least an order of magnitude~\cite{Armengaud:2014gea}.) While a kg-year exposure can cut into the QCD axion parameter space, stellar constraints remain stronger.  Superconductors will be a strong alternative, however, to model-dependent stellar constraints.

\subsection{Scalars}

We now consider scalar DM $X=\phi$ coupling to electrons via
\begin{equation}
	 {\cal L}\supset d_{\phi ee}\sqrt{4\pi}  \frac{m_e}{M_{pl}}\phi \bar e e\,,
\end{equation}
where we follow the normalization of Refs.~\cite{Adelberger:2003zx,Arvanitaki:2015iga}. Similar to the hidden photon and axion, the relic abundance of scalar DM can be set by a misalignment mechanism~\cite{Arvanitaki:2014faa}.

The dominant piece of the matrix-element-squared for the absorption of a scalar is
\begin{align}
	 |{\cal M}|^2 \approx    \frac{3}{\alpha} \left( \frac{ d_{\phi ee} m_e}{M_{pl} } \right)^2 \frac{\omega^2}{ |\vec Q|^2} |{\cal M}_\gamma|^2\,,
	\label{eq:Msquared_scalar}
\end{align}
and thus differs in $\vec Q$-dependence from the photon case, suppressed in comparison by $\omega^2/|\vec Q|^2$. Performing the integration in Eq.~\eqref{eq:response} and comparing with the photon rate in Eq.~\eqref{eq:rateQFT}, we thus obtain an $\omega$-dependent mapping from $\sigma_1$ to the scalar case. We arrive at a rate for scalar absorption of
\begin{align}\label{eq:Rscalar}
	R = \frac{1}{\rho} & \frac{\rho_X}{m_X}  \frac{3}{\alpha} \left( d_{\phi ee} \frac{m_e}{M_{pl}}  \right)^2 \sigma_1  \nonumber \\
	& \times \begin{cases}
 	\frac{5}{2}c_s^2 &,\ \  \omega < \omega_D \\
	\frac{5}{3} \frac{c_s^2 \omega^2}{\omega_D^2} \frac{\left(1 - \frac{3 \omega_D}{4 \omega} \right)}{\left(1 - \frac{5 \omega_D}{6 \omega} \right)}   &,  \ \ \omega > \omega_D
	\end{cases}
\end{align}
We use this result over the entire $\omega$ range shown, even though it does not account for interband transitions which are relevant above $\omega \gtrsim 0.5$~eV. Nevertheless, we expect Eq.~\eqref{eq:Rscalar} to be a reasonable proxy because the phase-space volume factor favors large phonon energies near the upper limit of $\omega_D$, where the suppression factor appearing in Eq.~\eqref{eq:Msquared_scalar} is well-captured by $\omega^2/| \vec Q|^2\sim \omega^2/| \vec Q_D|^2$ with $|\vec Q_D| = \omega_D/c_s$, up to an ${\cal O}(1)$ factor.

The projected sensitivity of a superconducting aluminum target for scalar DM absorption is presented in Fig.~\ref{fig:scalar}, for a kg-day (thin solid black curve) and kg-year (thick solid black curve) exposure. For comparison, we present the fifth-force constraints of Ref.~\cite{Adelberger:2003zx} (shaded blue and solid red curve), using the translation $|\alpha_{\rm mod}| = (d_{\phi e e} Q_e)^2$ where $Q_e \approx 1/4000$ is the fractional rest mass in electrons.  For masses above 0.1~eV, the derived constraints come from Casimir force experiments, and are not as rigorous.
We also plot HB cooling constraints (shaded purple), applying the limit 
$g_{\phi e e} \lsim 1.3 \times 10^{-14}$~\cite{Raffelt:1996wa,Grifols:1988fv} and setting $g_{\phi e e} = d_{\phi ee}\sqrt{4\pi}  \frac{m_e}{M_{pl}}$. %
Similar to the pseudoscalar case, the loop-induced coupling to photons
\begin{align}
	\frac{\alpha}{3 \sqrt{\pi} } \frac{d_{\phi ee}}{M_{pl} } \phi F_{\mu \nu} F^{\mu \nu}\,
	\label{eq:scalar_photon}
\end{align}
yields limits on $d_{\phi ee}$ from the $\phi \to \gamma \gamma$ decay time compared telescope searches~\cite{Grin:2006aw}, which we plot as well (dashed green outline). We find the superconducting detector gives the best sensitivity above 30 meV. Finally, we note the entire unexplored portion of the parameter space shown is technically natural, in that the $\phi$ mass-correction due to the electron coupling leads to $\delta m^2_\phi/m^2_\phi < 1$.

\section{Conclusions}\label{sec:conc}

We have explored the prospects of detecting ultralight DM, with mass in the meV to 10~eV range, via absorption in an aluminum superconductor. We find that even with modest exposure, the aluminum superconductor is particularly powerful for the case of hidden photon DM, easily superseding stellar constraints.  In the case of a light pseudoscalar, absorption on a superconducting target can also cut into the QCD axion parameter space.  Likewise, superconductors can probe scalar DM parameter space beyond constraints from stellar emission and fifth-force searches. Our results are summarized in Figs.~\ref{fig:sensitivityV},~\ref{fig:sensitivityA} and~\ref{fig:scalar}. Strikingly, the excellent reach of the superconducting targets is obtained despite the fact that the proposed detection method does not make use of DM coherence effects in the absorption process.

In fact, the DM mass range accessible to a superconducting absorber is exactly the mass range where the behavior of light bosonic DM transitions to that of a classical field, at masses of an eV. For masses well below this range, experimental techniques can rely on the coherence of the DM field to probe extremely small couplings. Our method, however, does not require a long coherence time of the DM field.  The DM signal is a single-particle mono-energetic absorption, which takes advantage of the superconductor sensitivity to an electronic excitation with energy as low as $\sim$~meV.

In a future publication, we will present the sensitivity of semiconducting targets to DM with masses above an eV via a similar absorption process, where we expect excellent reach~\cite{semiconductors}.

{\em Acknowledgments.}
We thank John Clarke, Adolfo Grushin, Roni Ilan, Maxim Pospelov, and Jakub Scholtz for useful discussions.  YH and KZ thank Matt Pyle and Yue Zhao for collaboration on our earlier work establishing superconductors as viable DM detectors.  We thank Yue Zhao for comments on the manuscript. YH is supported by the U.S. National Science Foundation under Grant No. PHY-1002399. TL is supported by DOE contract DE-AC02-05CH11231 and NSF grant PHY-1316783. KZ is supported by the DoE under contract DE-AC02-05CH11231.

\appendix
%

\section{Electrodynamics of solids}\label{app:Pi}

For an isotropic medium, the dielectric constant $\hat \epsilon$ is related to the complex index of refraction $\tilde n$ and is given in terms of the conductivity $\hat \sigma$,
\begin{equation}
	\hat \epsilon = \tilde n^2 = 1 + \frac{i \hat \sigma}{\omega}\,,
\end{equation}
 where we assume Lorentz-Heaviside units. 
The conductivity is directly related to the in-medium polarization tensor $\Pi^{\mu \nu} = e^2 \langle J_{\rm EM}^{\mu \dagger}, J_{\rm EM}^\nu \rangle$,
\begin{align}
	 \Pi^{\mu \nu}(\vec q, \omega) = \Pi(\omega) \sum_{i = 1,2} \epsilon_i^{T \mu}\epsilon_i^{T * \nu} + \Pi(\omega) \epsilon^{L \mu} \epsilon^{L \nu}
\end{align}
where $\epsilon_L, \epsilon_T$ are longitudinal and transverse polarizations vectors.
As described in Section~5.2 and Appendix~A of Ref.~\cite{Hochberg:2015fth}, $\Pi^{\mu \nu}$ is related to the dielectric constant, and for a non-magnetic medium,
\begin{eqnarray}
(\omega^2-\vec q\,^2)(1-\tilde{n}^2) & = &  \Pi_L\,, \no\\
\omega^2(1-\tilde{n}^2) & = &  \Pi_T\,.
\label{PiLPiT}
\end{eqnarray}
In the local limit of $|\vec q| \ll \omega$ the longitudinal and transverse $\Pi_L$ and $\Pi_T$ can both be written as Eq.~\eqref{eq:pisig},
\begin{align}
	\Pi(\omega) \approx  -i \hat \sigma \omega\,.
\end{align}

\bibliography{superconducting}

\end{document}